# Characterization of atmospheric pressure plasma plume

G.Veda Prakash*, Narayan Behera, Kiran Patel and Ajai Kumar
Institute for Plasma Research, HBNI, Gandhinagar, Gujarat - 382428, India
*Email: prakashgveda@gmail.com

**Abstract**

In order to optimize the parameters of the plasma plume for atmospheric pressure plasma applications such as biological and industrial applications, it is highly necessary to thoroughly understand its characteristics. In this paper, various diagnostic techniques are discussed to characterize the atmospheric pressure plasma plume. The major emphasis of this work is to utilize possibly simple methods, low complexity in post data analysis and obtain in-situ information. The parameters of the plasma plume such as spatial variation of gas temperature, electron excitation temperature, plume velocity, plume current, electron density are experimentally measured. In addition, plasma discharge behavior and plume formation with various gas flow rates and applied voltage are studied. The characterization of the above parameters is carried out by using electrical diagnostics (voltage probe and current sensor), Optical diagnostics (optical emission spectroscopy (OES), fast imaging method using ICCD camera), and thermocouple. The plasma plume/gas temperature along the plume length is measured by using a simple K-type thermocouple and is found to be in the range of (42.5 - 24.5)°C. The OES is used to estimate the electron excitation temperature of plasma plume and to identify the key reactive species (OH, NO, $N_2^+$, O, etc.) present in the plume. ICCD camera is utilized to capture the temporal evolution of plasma and to estimate the plume velocity along the plume length. The current transformer is utilized to measure the plume current along the plume length. By using the velocity and current information, the plasma density is evaluated, and obtained values are in the range of (0.069-5.96) × $10^{12}$ cm$^{-3}$. The above range of parameters and tunable nature of our designed plasma source will be useful for the development and optimization of plasma sources for various applications.

## 1. Introduction

In recent years, the demand for non-thermal atmospheric pressure plasma is increasing as it has proven promising for several applications in biological and industrial areas [1] [2] [3]. The plasma generated by using gases such as He, Argon, air, etc. are sources of reactive species, which can take part in various chemical reactions. Many researchers have presented various kinds of plasma devices based on applied voltage and method of the plasma generation such as dielectric barrier plasma pencil, double and single electrode plasma jet/plume driven by high-frequency RF, Pulsed DC power supplies [4] [5]. These devices have characteristics to produce plasma in the discharge area and deliver it into the ambient air. The non-equilibrium nature of the plasma makes it different from thermal plasma and makes it more complex in the aspect of the physics and chemistry involved in it [2]. Its non-thermal nature (heavy particle temperature remains close to the room temperature) makes it suitable for biomedical applications like wound healing, tissue sterilization, blood coagulation, electrosurgery, etc., as it avoids the heat effect or damage of living cells. One of the advantages of the generation of this plasma is it does not require the expensive and complex vacuum operation methods.

The parameters of plasma plume such as electron excitation temperature ($T_{exc}$) and gas temperature ($T_g$), electron density ($n_e$), and plume reactive species, plume velocity, etc. must be needed to understand to utilize it for biological and industrial applications. In the past, several experimental studies have taken place in this direction using various diagnostics such as electrical diagnostics, optical emission spectra (OES) and time-resolved fast imaging using an intensified charge-coupled device (ICCD) images [6] [7], etc. The relationship between the plasma plume length, charge and the applied voltage waveform under various gas flow rates have been investigated by [8] [9] and it is determined that the plume length increases with increasing plasma plume charge in laminar flow and follows the linear relationship. Here, the plume charge serves as the energy source so that the electric charge travels the plasma. Further, the information on plasma parameters such as electron temperature, density is estimated using optical emission spectroscopy (OES) techniques, which is non-invasive and required only moderate spectroscopic equipment [10]. An overview of the range of electron densities and gas temperatures that are encountered in typical atmospheric pressure plasmas are described in [9]. The electron excitation temperature in helium plasma inside the quartz tube was estimated using a Boltzmann plot method. Electron density has been estimated using optical emission technique [11] [12] [13], and charge density [14] methods and it is found to be in the range of around $10^{18}$ m$^{-3}$ at plasma discharge. In atmospheric pressure plasmas, the plume temperature normally estimated by van der Waals broadening [12] [15] [16] a spectral fitting of OH and $N_2$ second positive [17] [18] [19] spectra. The measured temperatures found to be in the range from 300 K- 325 K. These ranges of temperatures are suitable for biological (bacterial and fungal killing) and industrial applications, as it will kill the foreign bodies (bacteria, virus) and will not harm the living cells. One of the crucial parameters need to be understood in the plasma jet is its plume dynamics in the ambient air [20]. In the recent past, several researchers have studied the time evolution of the spatiotemporal structure of plasma into the ambient air as bullets or packets using fast imaging ICCD cameras [6]. To the naked eye, the plasma jet appears to be a continuous flow, however, in reality, it was observed to be constituted of discrete ionization wavefronts moving at a velocity much higher (in the order of 5-20 km/s) than the gas velocity [21] [3] [22]. In addition, it was deduced that the APPJ is mainly an electrical phenomenon and not a flow-related one.

In this work, the motivation is to utilize atmospheric plasma plume for biological (medical and agriculture) and industrial applications. For this purpose, the plasma plume needed to be thoroughly characterized (i.e. an investigation of the parameters of plasma and plume dynamics). To fulfill this goal, the diagnostic tools of quality such as low complexity in post data analysis, easy to handle, and obtain in-situ information are utilized. The plasma plume dynamics with various gas flow rates are obtained using a digital camera and electrical diagnostics. OES is used to measure the electron excitation temperature ($T_{exc}$) at various positions of the plume and to identify the reactive species present in the plume. A simple K- type thermocouple is utilized to estimate the plasma temperature along the plume length and found to be 42.5°C – 24.5°C in range, which is desired for several biological applications [23] [24] [25] [26] [27] [28] [29]. The advantage of using this simple and direct Thermocouple method over the techniques such as van der Waals broadening, spectral lines fitting is it avoids the significant post-processing to obtain the temperature value. In addition, this method is inexpensive and provides an in-situ reading of plasma temperature [4] [30], which is essential for field use. Further, in order to understand the plasma plume dynamics such as the plasma propagation, geometrical structural variation of the flowing plasma, temporal evolution into the ambient air, are obtained using ICCD images captured along the plume length. The plasma plume's current profile along the length of the plume is obtained using the current transformer (CT). By utilizing, the drift velocity (obtained from time-resolved images) and plasma current (current transfer) information, plasma density along the plume length is estimated.

The present paper is arranged as follows. In section 2, the experiemtnal scheme and diagnostics arrangement for plasma plume characterization is described. Results obtained using various electrical and optical diagnostics are discussed in section 3. The conclusion of the observation is presented in section 4.

## 2. Experimental scheme

The experimental scheme has been presented in our previous report [38]. Only a brief summary, which is important for the present study, is presented here. The schematic view of the experimental setup and diagnostic arrangement for the present work is shown in Figure 1(a). A sinusoidal high voltage (4 kV$_{p-p}$), 33kHz is used to produce a plasma plume of ~ 4 cm. Pyrex glass tube of 4 mm inner diameter and 1 mm thickness is used in this work. For plasma generation, the electrodes are arranged in a cross-field configuration as shown in Figure 1(a). Helium (He) gas of 99.99% purity is used as an active gas for plasma generation. The photograph of visible plasma plume is shown in Figure 1(b). In the present manuscript, the operating conditions i.e. applied voltage is 4 kV$_{p-p}$, and the Helium gas flow rate is 11 lpm which are fixed for entire characterization. The details of the operating condition found are discussed in section 3.1. In this work, the special emphasis is given on to utilize diagnostics, which are simple to use and provide in-situ information for characterizing the parameters of plasma. The parameters such as gas temperature, plasma density, plasma velocity play a key role in various biological and industrial applications. The tools used to characterize plasma plume are (i) high voltage probe for discharge voltage profile, (ii) current transformer (CT) for discharge current profile and plume current measurement, (iii) spectrometer for electron excitation temperature and active species identification, (iv) thermocouple for gas temperature measurement, and (v) ICCD camera for plume propagation velocity and plume dynamics. These tools are arranged as shown in Figure 1 (a). The voltage and current profiles from the voltage probe and CT respectively are captured in the Oscilloscope. A collecting lens along with optical fiber cable is utilized to acquire the emission profile of the plasma discharge, which is recorded in a computer. The position of collecting lens is placed at a distance from the plume such that it does not disturb the plasma and also acquire the emission profile without loss of information The collecting lens is fixed on the jack that enabled to scan the various positions by varying the jack height. In order to examine the plasma temperature at a specific location along the plume length, the extended length of the thermocouple cord is placed on the jack whereas its tip is ensured in contact with the plasma to measure the plasma temperature. The ICCD camera along with the focal lens is used to capture the plasma plume bullets propagation and geometrical dimensions of the plume. The lens focus is adjusted such that it covers the entire length of the plume from glass nozzle to plume tip and without losing any light information. The technical specifications of diagnostics have discoursed below and details are discussed at respective sections.

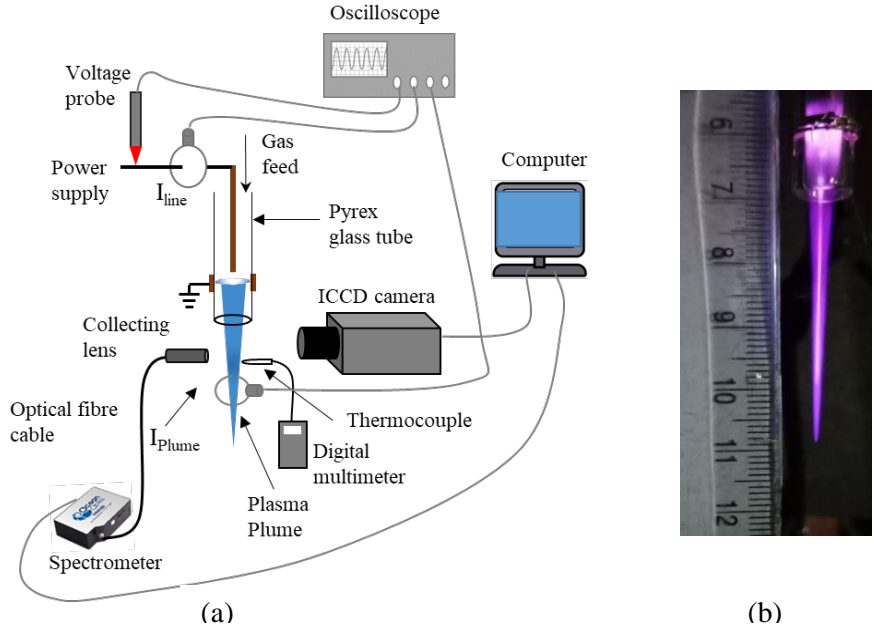

*Figure 1. (Color online) (a) Schematic showing various diagnostics arrangement for characterization of plasma plume, (b) Typical photograph of the visible plasma plume in the ambient air.*

In the present work, Tektronix P6015 A (75 MHz bandwidth) voltage probe is used to measure the applied voltage. Two current transformers (FCT-016-20:1, Bergoz instrumentations, 1.25V/A) are utilized, one is for measuring line current ($I_{line}$) which is equal to the total current supplied by the power source and other is to estimate plasma plume current ($I_{Plume}$). The applied voltage and current waveforms are recorded using a digital storage oscilloscope (DSO -X 2024A, Agilent Technologies, 200MHz bandwidth, 2 GSa/s) to analyze the discharge parameters with various gas flow rates. The optical emission spectra is recorded using Ocean optics HR-4000 spectrometer to estimate the electron excitation temperature and to identify the reactive species (OH, NO, $N_2^+$, O, etc.). A K-type thermocouple with a digital multimeter (Fluke 287) has been used to measure the gas temperature along the length of the plasma plume. Time-Resolved fast images of flowing plasma plume are captured using an intensified charge-coupled device (ICCD, 4 Picos, Stanford computer optics) camera fitted with an AVENIR lens (25 mm, F 1.4). The digital image processing technique is used to obtain the velocity of plasma plume and the dimensions of the plume. Data obtained from ICCD and CT are utilized for density evaluation along the plume length.

## 3. Results and discussion

### 3.1 Operating conditions

Figure 2 shows a typical applied voltage and discharge current (line current) profiles during full-length (~ 4 cm) plasma plume formation in the ambient air. The discharge occurs in each half cycle of the applied voltage, which can be further observed from the sharp discharge peaks present in the current waveform. The maximum applied voltage in the present experiments is 4 $kV_{p-p}$ and the discharge current is 10 $mA_{p-p}$.

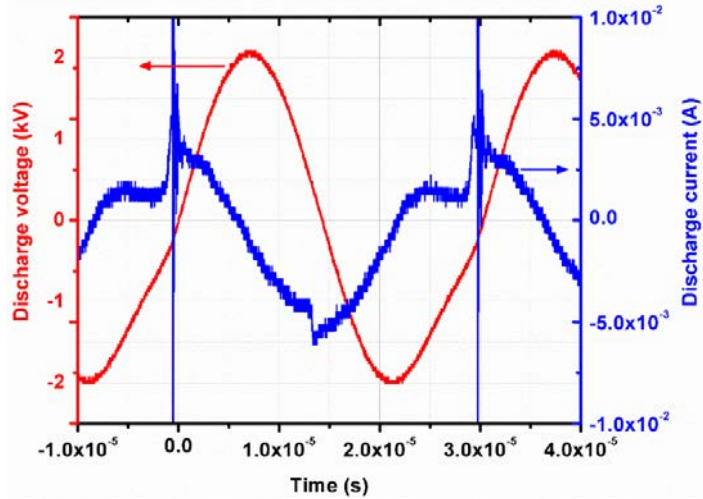

*Figure 2. (Colour online) Typical applied voltage, discharge current waveforms for ~ 4 cm plasma plume formation.*

The dynamics of the plume formation in the ambient air with reference to the incremental gas flow are analyzed using photographs taken by a digital camera. The plot for plume length vs gas flow rate is presented in Figure 3(a) where the applied voltage is fixed at $4kV_{p-p}$. It shows that until 2 lpm gas flow, there is no plume formation in the air. At 4 lpm of gas flow plume reaches around 0.7 cm, and it increases up to 3.3 cm at 8.5 lpm and finally, it approaches 4 cm at 11 lpm where the maximum plume attained. The plume gets disturbed with a further increment of gas flow. This information shows the maximum and steady plasma plume is dependent on the optimized gas flow rate.

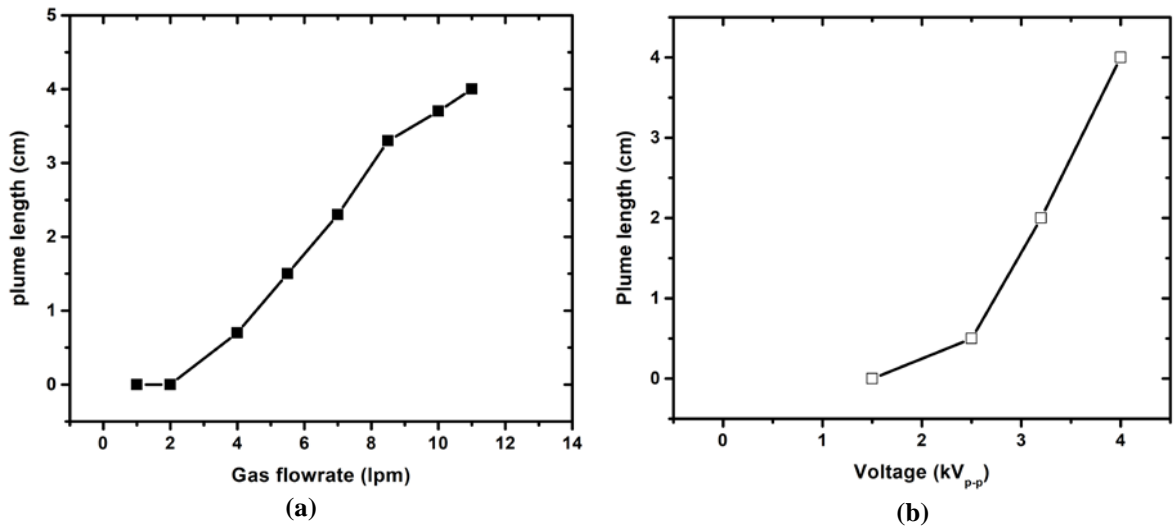

*Figure 3. Plasma plume length in the ambient air with increment in (a) gas flow rate, with fixed voltage (b) applied voltage with vs plume length.*

Further, the plume dynamics with applied voltage are shown in Figure 3 (b). To initiate discharge, the minimum voltage required is 1.5 kVp-p with the gas flow around 1 lpm where it observed as a discharge channel. As the voltage increases to 2.5 kV, the discharge further increases as multiple discharge channels and plasma forms up to glass nozzle from the origin of the discharge. As the gas flow increases 7 lpm with applied voltage 2.5 kVp-p, the plasma plume appears as a narrow plume with a low intense channel of 5 mm length into the ambient air. At around 3.2 kVp-p, the plasma forms as a very bright channel and achieved a better plume length of more than 2 cm into the ambient air with a flow rate around 10 lpm. The maximum plume length achieved is 4 cm at 4 $kV_{p-p}$ with a gas flow rate of around 11 lpm. These are the operating conditions for the plasma plume jet. At reduced gas flow rates with variable voltage, the observed plume has a thin channel with less intensity and length. The above discharge properties show that the maximum plasma plume length is achieved at optimum operating voltage and gas flow rate.

## 3.2 Optical emission studies

To determine the plasma composition and electron excitation temperature ($T_{exc}$), an optical emission spectroscopy method is utilized. The full range spectrum is recorded by using an Ocean optics HR-4000 spectrometer and the emitted lines were observed in the range of 200 to 900 nm. The spectral light is viewed normal to the plume expansion and fed at the entrance slit of the spectrometer through the optical fiber. The $T_{exc}$ and reactive species are evaluated using the intensity of atomic emission lines obtained from the spectrometer. $T_{exc}$ is determined from the spectral data using the Boltzmann plot method [32]. In spectra, the excited and ionized states of oxygen, nitrogen, and hydroxyl radicals were observed along with He lines, The spectrum with prominent lines identified (which are also the important reactive species required for several biological applications) is shown in Figure 4. The presence of OH, NO, $N_2^+$, O, etc. is identified in accordance with the NIST database [39], which affirms the suitability of our plasma plume for biological and industrial applications [20]. The electron excitation temperature ($T_{exc}$) obtained using the above observed He lines by Boltzmann plot method. $T_{exc}$ is estimated by using the four prominent He lines 587.6, 667.8, 706.6 and 728.1 nm and by the relation (1).

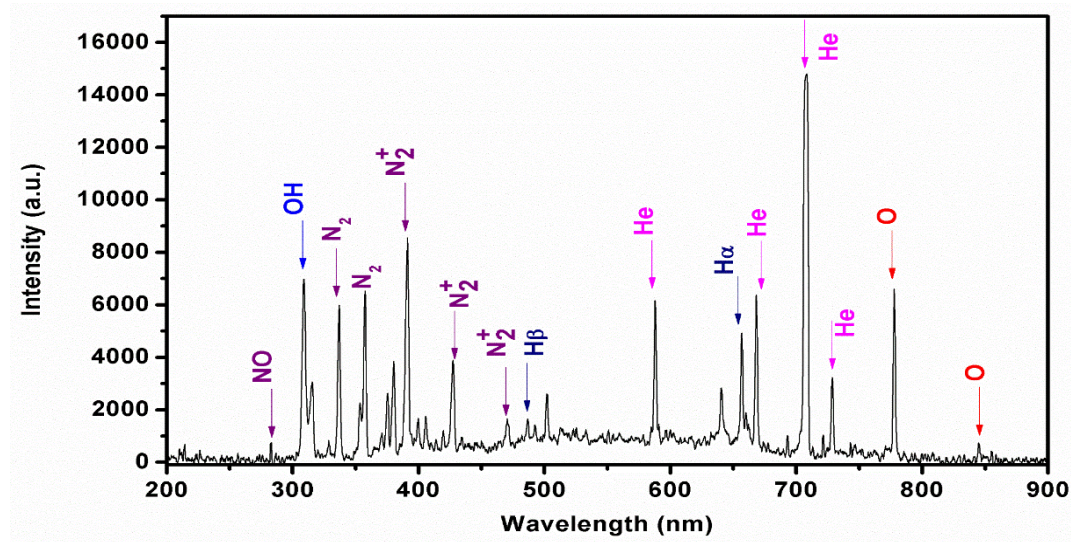

*Figure 4. (Colour online) Optical emission spectra of plume showing important active species in 200-900 nm wavelength range.*

$$\ln\left(\frac{I_n \lambda_n}{g_n A_n}\right) = -\frac{E_n}{T_e} + \ln\left(\frac{Nhc}{U}\right) \quad (1)^\dagger$$

where $I_n$ is the intensity of the spectral line of wavelength $\lambda_n$, $g_n$ is the statistical weight factor, $A_n$ is the transition probability, $T_{exc}$ is the excited electron temperature, $E_n$ is the energy of the excited states of the spectral line, N is the molecular density, h is the Planck's constant, c is the speed of light, and U is the partition function. The intensity values of these spectral lines $I_n$ are obtained from the observed spectra. The values of $E_n$, $g_n$, and $A_n$ for the selected lines are taken from the NIST atomic spectra database.

---

†There was a typo in the earlier version (G. Veda Prakash et al., arXiv preprint arXiv:1912.03691 (2019).).

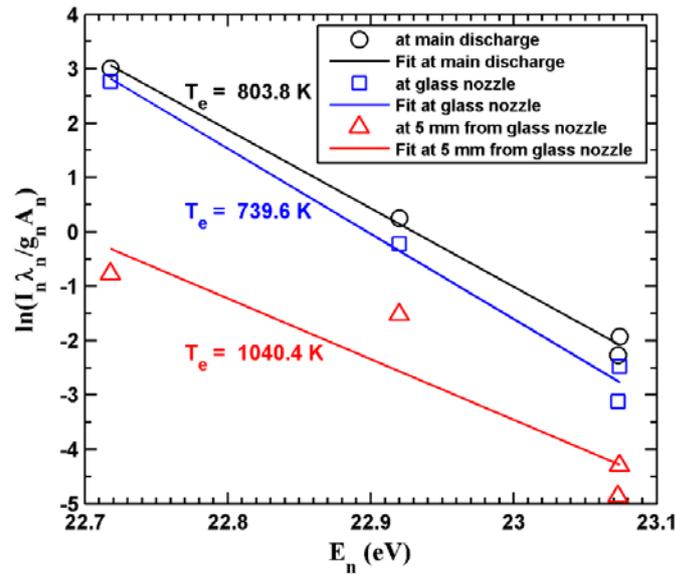

*Figure 5. (Color online) Boltzmann plot fitting for estimation of Electron excitation temperature at main discharge, glass nozzle, at 5 mm from the glass nozzle.*

The electron excitation temperature ($T_{exc}$) obtained at three different positions are shown in Figure 5. The temperature at the main discharge is found to be 804 K, and at the nozzle is around 740 K. Temperature towards the plume tip expected to decrease. However, the measurements obtained at 5 mm below the orifice i.e. in ambient air found to be 1040 K, which is believed to be overestimated value due to collisions between the helium gas and air [33].

### 3.3 Gas temperature from thermocouple

In order to estimate the plasma/gas temperature, the thermocouple measurements are performed along the plume length [30]. Standard K-type thermocouple connected to a digital multimeter (Fluke 287 model) is used for the measurement of the plasma plume temperature. The arrangement of thermocouple and technical specifications is discussed in the diagnostics arrangement section. The advantage of using thermocouple is discussed in the section-1. The measured plasma temperature is (42.5-24.5) ℃ in range along the plume length which is shown in Figure 6, which found to be an acceptable value for several medical and industrial applications [10].

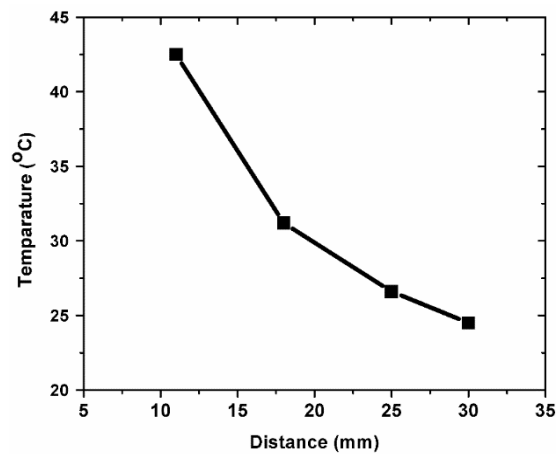

*Figure 6. Variation of plasma plume temperature along the length of the plume measured using K-type thermocouple. Error ± 2%.*

### 3.4 Plasma plume velocity

To develop an understanding of plasma plume dynamics such as plume velocity, experiments are performed using the ICCD (4 Picos, Stanford Computer Optics) camera by varying the delay time from 500-6000 ns and exposure time of 500 ns. Time-resolved images of visible plasma plume provide detailed information related to the hydrodynamic and geometrical structure of flowing plasma plume. The camera and the plasma voltage source are synchronized with a signal generator. A delay time of 500 ns increased for succeeding images of the plume. The details of the ICCD images and plasma plume dynamics are discussed in ref [38]. From the ICCD images, the evaluated plume bullet velocity at the glass nozzle is ~ 22.46 km/s and it decreased to 1.38 km/s as shown in Figure 7. The major reduction of plume velocity from the glass nozzle to the plume tip is due to ambient air drag force experience during the interaction of two fluids, i.e. plasma plume and ambient air. Further details of plume velocity and its geometrical and hydrodynamic properties are discussed in ref [38].

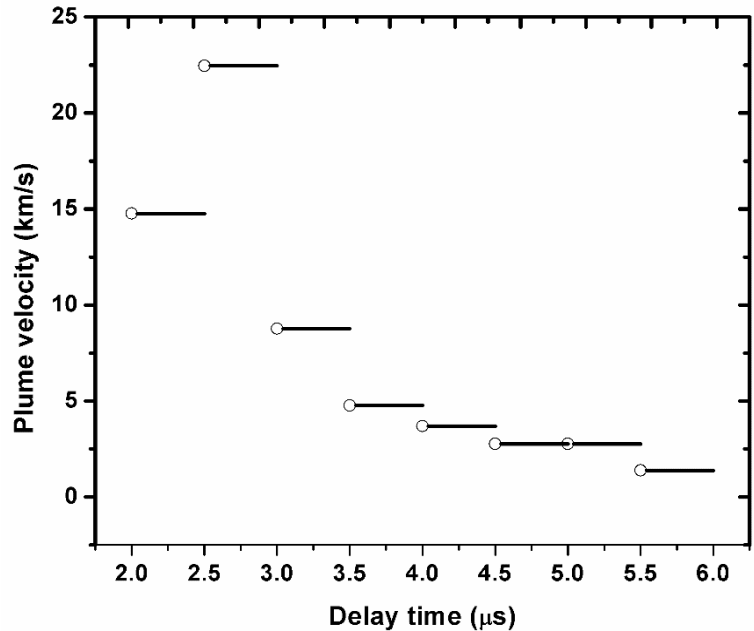

*Figure 7. Plasma bullet velocity profile with delay time.*

The measured velocity is reducing from the glass nozzle to the plume tip due to the interaction of two fluids. The variation pattern of velocity follows a similar pattern as reported in ref [37]

## 3.5 Plasma jet current

In order to understand the plume charge variation, in-situ measurement of plasma plume current is carried out using the Current Transformer (CT), the arrangement of which is shown in Figure 1. The plasma plume is released into the ambient air from the glass nozzle. The CT has an inner diameter of 10mm, and the plume has a maximum diameter of 3.3mm. Hence, the plume never affected the arrangement of CT along the length of the plume. In this work, the direct measurement of plume current has been carried out at three different locations along the plume length. The plume current profiles (3-5 µs pulse width) at the various position along the plume with reference to the power line current (discharge current) and applied voltage are shown in Figure 8 [7].

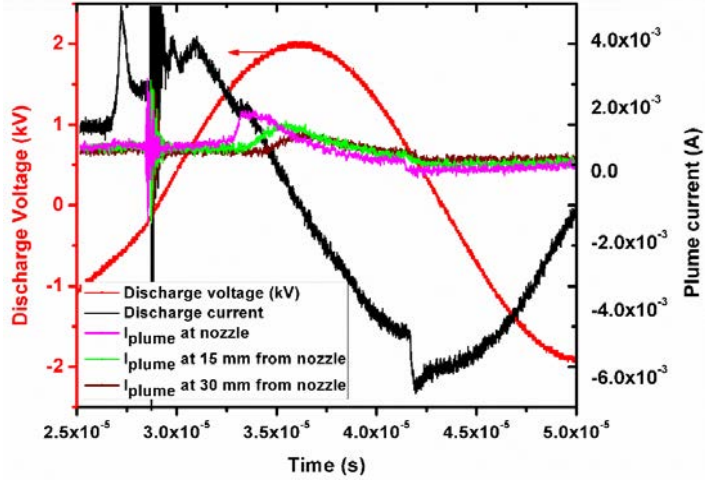

*Figure 8. (Colour online) Plasma plume current along the plume length.*

### 3.6 Measurement of plasma density along the plume length

From the previous sections, it is found that the plume dynamics (i.e. velocity, current, etc.) found to be decrement from glass nozzle to plume tip. Therefore, it will be interesting to find the variation in plasma density (hence pressure, chemical reactions, treatment duration), which will be useful for several applications. Plasma plume captured with 20 µs exposure time, shown in Figure 9, from which the plume cross-sectional area (S) is found to be 8.55, 4.15, 0.95 mm$^2$ at 10, 25 and 35 mm location from the main discharge. This variety of areas of cross-sections can be utilized for several applications depends on the area of treatment required (for ex: eye infection required very small area of cross-section of the plume, wounds may require large area). Hence, here the plasma density measurement is carried out along the plume length. As reported previously [38], in a weakly ionized plasma the plasma density is estimated using relation.

$$n_{plume} = \frac{I_{plume}}{ev_{dri}S} \qquad (2)$$

where $e$ is the elementary charge ($e$ = 1.609× 10$^{-19}$ C), $I_{plume}$ is plasma plume current, $v_{drift}$ is plume drift velocity, $S$ area of cross-section. $I_{plume}$ obtained from CT and $v_{dirft}$, and $S$ are from ICCD camera images. The calculated charge density for three different positions of the plume i.e., at 10, 25, 35 mm respectively from the main discharge is (0.0692, 0.13068, and 5.96) × 10$^{12}$ cm$^{-3}$ respectively which are in the limits of earlier reported values [36]. Further detailed discussion on plasma density along the length is presented in ref [38].

The above characterization of plasma plume can be summarised as follows. The optical emission spectra is used to identify the prominent species present in the plasma plume, which are essential for various biomedical applications [1-3] and electron excitation temperature. The gas temperature obtained from the thermocouple shows the suitability of the plasma plume for biomedical applications, as the temperature is around room temperature which does not cause thermal damage. The plasma bullet velocity obtained from the ICCD camera is useful to understand the impact it creates on the sensitive targets [1]. The plasma density value indicates the presence of energized electrons in the plume, which involves in ionization process and hence generate the active species required for various biomedical applications [1-3]. The details of the results are given in Table 1.

Table 1. Various diagnostics used for characterization of the atmospheric pressure plasma plume

| S. No. | Diagnostic tools | Measured parameters | Results |
| --- | --- | --- | --- |
| 1 | Optical emission spectroscopy | Identification of active species | OH, NO, $N_2^+$, O, etc |
| | | Electron excitation temperature | (804 – 1040) K |
| 2 | Thermocouple | Gas temperature | (42.5 – 25) °C |
| 3 | ICCD camera | Plume velocity | (22.46 – 1.38) km/s |
| 4 | ICCD camera + Current transformer | Plasma density | (0.0692 - 5.96) $\times 10^{12}$ cm$^{-3}$ |

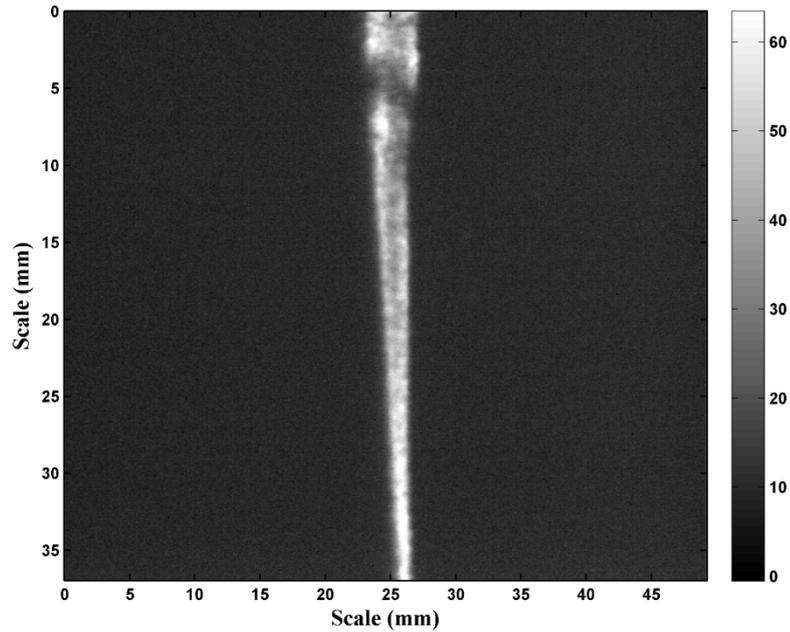

*Figure 9. Plasma plume captured with 20μs exposure time, showing geometry of plume.*

These parameters i.e. velocity and plasma density information along the plume length helps to decide the duration of the treatment and also helps to estimate the impact due to high electron density at the plume tip for very sensitive targets where the area of cross-section is highly critical such as treatment for eye-related infections and other *in vivo* applications.

### 4. Colnclusion

In summary, we have presented the non-thermal atmospheric pressure plume characterization for various biological and industrial applications. By using high voltage (4 kV p-p and 33 kHz frequency) source, a low-temperature atmospheric pressure plasma plume of length ~ 4 cm has been produced. The discharge ignition and plume formation with reference to incremental gas flow rate and applied voltage demonstrated the optimization between gas flow and voltage, which enables us to achieve the steady plasma plume as required for biological applications. The electron excitation temperature from main discharge to glass nozzle found to be 803.8 K, and 739.6 K respectively. The key reactive species such as O, OH, $N_2^+$, etc., present in the plume affirms the suitability of our plasma for several medical applications such as cancer, etc. The real-time plume dynamics obtained from the ICCD camera provides the plume initial velocity ~22.46 km/s and rapidly slows down to a velocity of 1.38 km/s due to the opposition from ambient drag. The plume current measurements along the plasma plume length measured are (1.27-2) mA. By utilizing the plume velocity and current information, the plasma density along the plume length is evaluated and it found to be in the range of (0.069-5.96) $\times 10^{12}$ cm$^{-3}$. Further optimization of the above parameters of the plasma plume is highly required for any individual applications. The above mentioned diagnostic techniques and characterization of plasma plume have very significance to a large number of biological applications.